\documentclass[manuscript]{aastex}

\shorttitle{Polarimetry Observations}
\shortauthors{Aller et al.}

\begin{document}


\title{Constraining  the Physical Conditions in the Jets of $\gamma$-ray Flaring Blazars using Centimeter-Band Polarimetry and Radiative Transfer Simulations.
 I. Data and Models for 0420-014, OJ 287, \& 1156+295}


\author{M. F. Aller, P. A. Hughes, H. D. Aller, and G. E. Latimer\altaffilmark{1}}
\affil{Department of Astronomy, University of Michigan, Ann Arbor, MI 48109-1042}
\email{mfa@umich.edu}

\and

\author{T. Hovatta\altaffilmark{2}}
\affil{Cahill Center for Astronomy and Astrophysics, California Institute of Technology, 1200 E. California Blvd, Pasadena CA 91125}

\altaffiltext{1} {deceased}
\altaffiltext{2} {Aalto University Mets\"ahovi Radio Observatory, Mets\"ahovintie 114, 02540 Kylm\"al\"a, Finland}



\begin{abstract}
To investigate parsec-scale jet flow conditions during GeV $\gamma$-ray flares detected by the {\it Fermi} Large Angle Telescope, we obtained centimeter-band total flux density and linear polarization monitoring observations from 2009.5 through 2012.5 with the 26-meter Michigan radio telescope for a sample of core-dominated blazars. We use these data to constrain radiative transfer simulations incorporating propagating shocks oriented at an arbitrary angle to the flow direction  in order to set limits on the jet flow and shock parameters during flares temporally associated with $\gamma$-ray flares in 0420-014, OJ~287, and 1156+295; these AGN exhibited the expected signature of shocks in the linear polarization data. Both the number of shocks comprising an individual radio outburst (3-4) and the range of the compression ratios of the individual shocks (0.5-0.8) are similar in all three sources; the shocks are found to be forward-moving with respect to the flow. While simulations incorporating transverse shocks provide good fits for 0420-014 and 1156+295, oblique shocks are required for modeling the OJ~287 outburst, and an unusually-low value of the low energy cutoff of the radiating particles' energy distribution is also identified.  Our derived viewing angles and shock speeds are consistent with independent VLBA results. While a random component dominates the jet magnetic field, as evidenced by the low fractional linear polarization, to reproduce the observed spectral character requires that  a significant fraction of the magnetic field energy is in an ordered axial component.  Both straight and low pitch angle helical field lines are viable scenarios. 
 \end{abstract}





\section{Introduction}
 Blazars are an active galactic nuclei (AGN) class comprised of BL Lacertae objects (BL Lacs) and flat spectrum radio quasars (FSRQs) based on their optical properties. The emission from these sources is characterized by high levels of variability (flares) across the spectrum from the radio band to the $\gamma$-ray spectral region, the presence of variable linearly-polarized flux in both the optical and radio bands, and beamed non-thermal emission. A complementary classification system based on the non-thermal emission from AGN, including blazars, uses the location of the peak of the synchrotron bump in the broadband Spectral Energy Distribution (SED) as a class delineator; these SED classes are designated low-synchrotron-peaked (LSP), intermediate-synchrotron-peaked (ISP), and high-synchrotron-peaked (HSP), with peaks below 10$^{14}$ Hz, between 10$^{14}$ and 10$^{15}$ Hz, and higher than 10$^{15}$ Hz respectively. This peak location is related to the maximum energy of the accelerated electrons and is particularly relevant for studies of $\gamma$-ray-bright sources \citep{abd10b}. The non-thermal emission arises in highly-collimated, relativistic outflows (jets) oriented at low angles to the observer's line of sight, and emanating from a central supermassive black hole and associated accretion disk. 

The detection of GeV $\gamma$-ray emission from $\approx$ 70 blazars with high significance by EGRET aboard the Compton Gamma-Ray Experiment in the 1990s \citep{har99} first identified the importance of the high energy emission in the energy budget of these sources, and work during this era provided the first evidence supporting a scenario in which the high energy emission arises within the parsec scale region of the jet \citep{val95}. Radio-band data played an important role in establishing the presence of correlated activity during some events, consistent with a scenario where a common disturbance produces the broadband activity \citep{val96}. The identification of the ejection of new emission features from the inner jet core associated with the $\gamma$-ray flaring \citep{jor01} provided further support for a causal relation between the radio-band outbursts and the $\gamma$-ray flares.  However, attempts to fully understand the origin of the high energy emission were hampered by both the erratic data sampling resulting from the use of a pointed mode of observation and by the relatively low sensitivity of EGRET. With the launch of $\it Fermi$ in mid-2008 yielding consistent sky coverage and a high temporal sampling rate provided by sky scans every three hours, it was expected that definitive answers concerning the origin (site and physical processes) of the  GeV $\gamma$-ray emission would be obtained. The 2FGL catalogue, based on only the first two years of the operation of $\it Fermi$ in survey mode, contains detections for nearly 1000 AGN, primarily of the  blazar class \citep{ack11,nol12} providing a plethora of data. However, in spite of these new measurements, and the wealth of contemporaneous broadband data from both ground-based and satellite instruments, basic questions have remained unanswered or subjects of controversy. These open questions include the source of the seed photons for the production of the $\gamma$-ray flares, identification of an acceleration process for producing the high energy particles, localization of the emission site (within the parsec-scale jet, or near the central supermassive black hole, or in more than one location), and the specification of the unique physical conditions within the parsec-scale jet (or elsewhere), which are required for the production of the $\gamma$-ray flares. These open questions motivated the work described here which uses radio-band polarimetry to probe the magnetic field structure in the emitting regions and to identify changes in the flow consistent with the passage of shocks temporally associated with the $\gamma$-ray activity.

The mechanism for producing the particle acceleration required for up-scattering lower energy ambient photons from synchrotron-emitting electrons to $\gamma$-ray energies in the relativistic jet remains an important subject of debate, and a variety of processes have been discussed in the literature to explain the $\gamma$-ray flares, including scenarios invoking shocks. Internal shocks develop naturally from instabilities within the relativistic flows \citep{hug05} and  have generally been accepted as the origin of major outbursts in both the radio band and the optical band since the 1980s \citep{hug85,mar85}. Recent proposed mechanisms for the generation of the $\gamma$-ray flares have included internal shocks produced when faster portions of the flow overtake slower ones or reconfinement shocks formed at distances where the supersonic jet is influenced by the external medium at parsec scales (see \citet{ack12} and references therein). Amplification of the magnetic field downstream of the shock with associated particle acceleration \citep{che13} is a common feature of shocks. However, while shocks have been proposed in several broadband studies of the $\it Fermi$ data during flares, supportive evidence based on polarimetry data has been limited due to the inadequate sampling cadence for tracing the expected changes, e.g. \citet{abd10a}; and quantitative tests based on theoretical studies of the emissivity incorporating shocks have been sparse \citep{all13b,weh12,mar14}.  

To identify the presence and possible role of shocks in the production of the $\gamma$-ray flares detected by {\it Fermi}, in 2009.5 we initiated a program to monitor a sample of blazars for total flux density and linear polarization with the University of Michigan 26-m radio telescope (UMRAO) at 3 centimeter-band wavelengths using an initial sample comprised of 33 AGN \citep{all09}. These observations continued through 2012.5 when the facility UMRAO was closed. Many of these AGN had previously been observed for up to 4 decades as part of the UMRAO variability program but generally with lower cadence before the commencement of the program described here. In this paper we present results for three blazars which exhibited large-amplitude outbursts in both the $\gamma$-ray and radio bands and evidence for the passage of shocks in the linear polarization data. We use the new data to constrain shock models developed in an earlier phase of the program  \citep{hug11} in order to identify the flow conditions in the parsec-scale jet during the $\gamma$-ray flaring. All three sources have flared in the $\gamma$-ray band since the launch of {\it Fermi}, and one of them has attained  maximum photon fluxes exceeding 10$^{-6}$ photons cm$^{-2}$sec$^{-1}$ based on weekly-averaged data. These blazars represent a variety of optical and SED classes and span a range of redshifts. While none are listed in TeVCAT\footnote{http://tevcat.uchicago.edu}, the online $\gamma$-ray catalogue of TeV-detected sources, all are included in the 1FHL catalogue \citep{ack13} based on their energies above 10 GeV; hence they represent GeV $\gamma$-ray sources in which  particle acceleration to high energies has occurred. The observing and reduction procedures, source selection, and variability properties are described in \S2; the model and model-fitting procedures are detailed in \S3;  the model fitting for the three sources is  described in \S4;  in \S5 we  discuss and compare these results; and in \S6 we  summarize our conclusions.

\section{The Observations}

\subsection{Centimeter-band Observational Procedures and Source Selection}

The UMRAO total flux density and linear polarization observations presented were obtained with the University of Michigan 26-m equatorially-mounted paraboloid as part of the University of Michigan AGN monitoring program \citep{all85a}.  At 14.5 GHz (2 cm), the primary frequency for the measurements since the launch of {\it Fermi}, the polarimeter consisted of dual, rotating, linearly-polarized feed horns,  symmetrically placed about the paraboloid's prime focus; these fed a broadband, uncooled HEMPT amplifier with a bandwidth of 1.68 GHz. At 8.0 GHz (4 cm) an uncooled, dual feed-horn beam-switching polarimeter system was employed using an on-on observing technique; the bandwidth at this operating frequency was 0.79 GHz.  At 4.8 GHz (6 cm) a single feed-horn system was employed with a central frequency of 4.80 GHz and a bandwidth of 0.68 GHz. A change of observing frequency required a horn change, and hence observations were generally carried out for 1-2 day time blocks at each frequency in a rotation. These observations were made within $\pm$2.5 hours of prime meridian passage in order to reduce the effect of position-dependent gain variations on the measurements. The adopted flux density scale is based on \citet{bar77} and uses Cas A as the primary standard.  Observations of nearby secondary flux density calibrators were interleaved with the observations of the target sources approximately every 1.5 to 3 hours to verify the stability of the antenna gain and to confirm the accuracy of the telescope pointing. The electric vector position angles (EVPAs) were calibrated using a source of polarized emission mounted at the vertex of the paraboloid; this assembly was surveyed at installation to an accuracy of 0.12$^{\circ}$. To verify the calibration and stability of the instrumental polarization, selected Galactic H~II regions were observed several times each day; these objects were assumed to be unpolarized. Each daily-averaged UMRAO observation was comprised of a series of 8 to 16 individual measurements obtained over a 25 to 45 minute time period. Because of the frequent observations of the calibration sources, the number of targets which could be monitored in a 24-hour run was limited to 20 to 24 sources. 

The initial source sample  of 33 sources for the project was selected on the basis of known or expected variability in both the $\gamma$-ray and the centimeter-band. In the radio-band the multi-decade, centimeter-band data from the UMRAO monitoring program provided a guide to the expected time scales of the variability, the peak amplitudes reached during flares, and the presence of ordered changes in the linear polarization data. Selecting sources with large-amplitude, resolved variations was crucial to the project goals since time-variability in the flux and linear polarization provide the constraints in the modeling. Sources with highly-significant EGRET detections were preferentially chosen initially, but new flaring sources were added based on the {\it Fermi} LAT measurements. In addition, the sample was restricted to sources in the MOJAVE (Monitoring of Jets in active galactic nuclei with VLBA Experiments\footnote{https://www.physics.purdue.edu/astro/mojave/}), 15.4 GHz imaging program (a frequency which is near to the UMRAO primary frequency and probes the same region of the jet), and priority was given to sources in the Boston University monitoring program at 43 GHz\footnote{http://www.bu.edu/blazars/VLBAproject.html}, a frequency which provides complementary information on the structure and structural changes in the inner jet region.  As the project progressed, the sample was reduced to 18-20 blazars in order to observe the current most-variable objects with increased cadence. 

\subsection{GeV-band Light Curves}

The $\gamma$-ray photon flux light curves for the three sources studied were produced in the 0.1-200 GeV band using $\it Fermi$-LAT data obtained during  August 2008 -- December 2012.  Weekly-binned photon fluxes were obtained using {\bf ScienceTools}--v9r27p1 and P7SOURCE\_V6 event selection. The LAT data were extracted within a 10$^\circ$ region of interest (ROI) centered upon the position of the target. These used an unbinned likelihood analysis (tool gtlike) to determine the photon fluxes by including in the model all of the sources within 15$^{\circ}$ of the target and by freezing the spectral index of all sources to the values in the 2FGL catalogue.

\subsection{Variability Histories: Importance and Results}

The past variability histories of the three sources based on the UMRAO monitoring data and their relation to EGRET detections are briefly reviewed below. These earlier measurements are useful for placing the variations observed during the operation of  {\it Fermi} in context, for estimating the duty cycle in the radio band, and for relating the onset of $\it Fermi$ flares to the radio-band data by identifying opacity effects. Synchrotron self-absorption can produce time delays between the peaks of the flares in the $\gamma$-ray and the centimeter bands ranging from zero to several months \citep{pus10}.

\subsection{Long-term Variability in 0420-014}

\begin{figure}
\begin{center}
\includegraphics[scale=0.50]{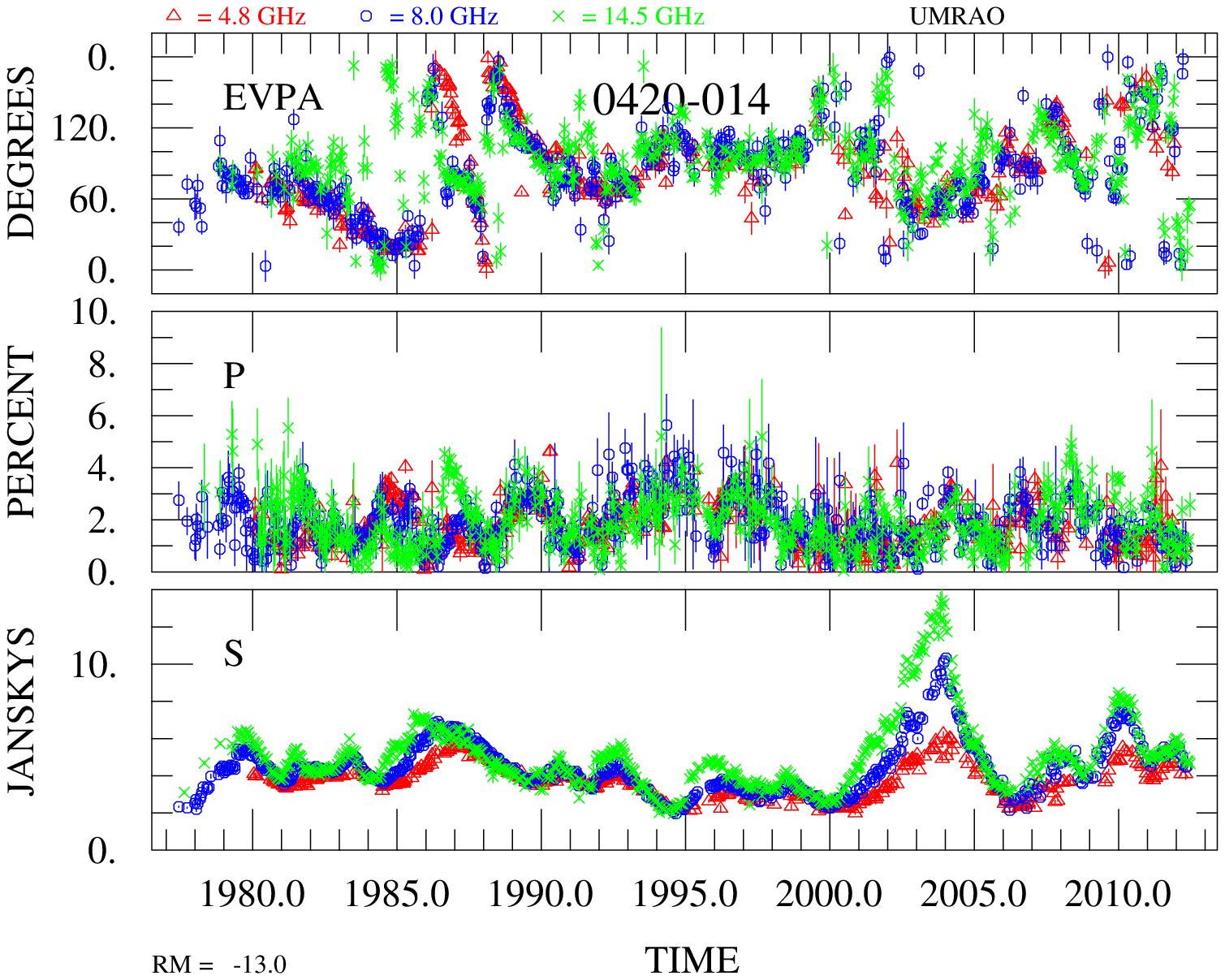}
\includegraphics[scale=0.40]{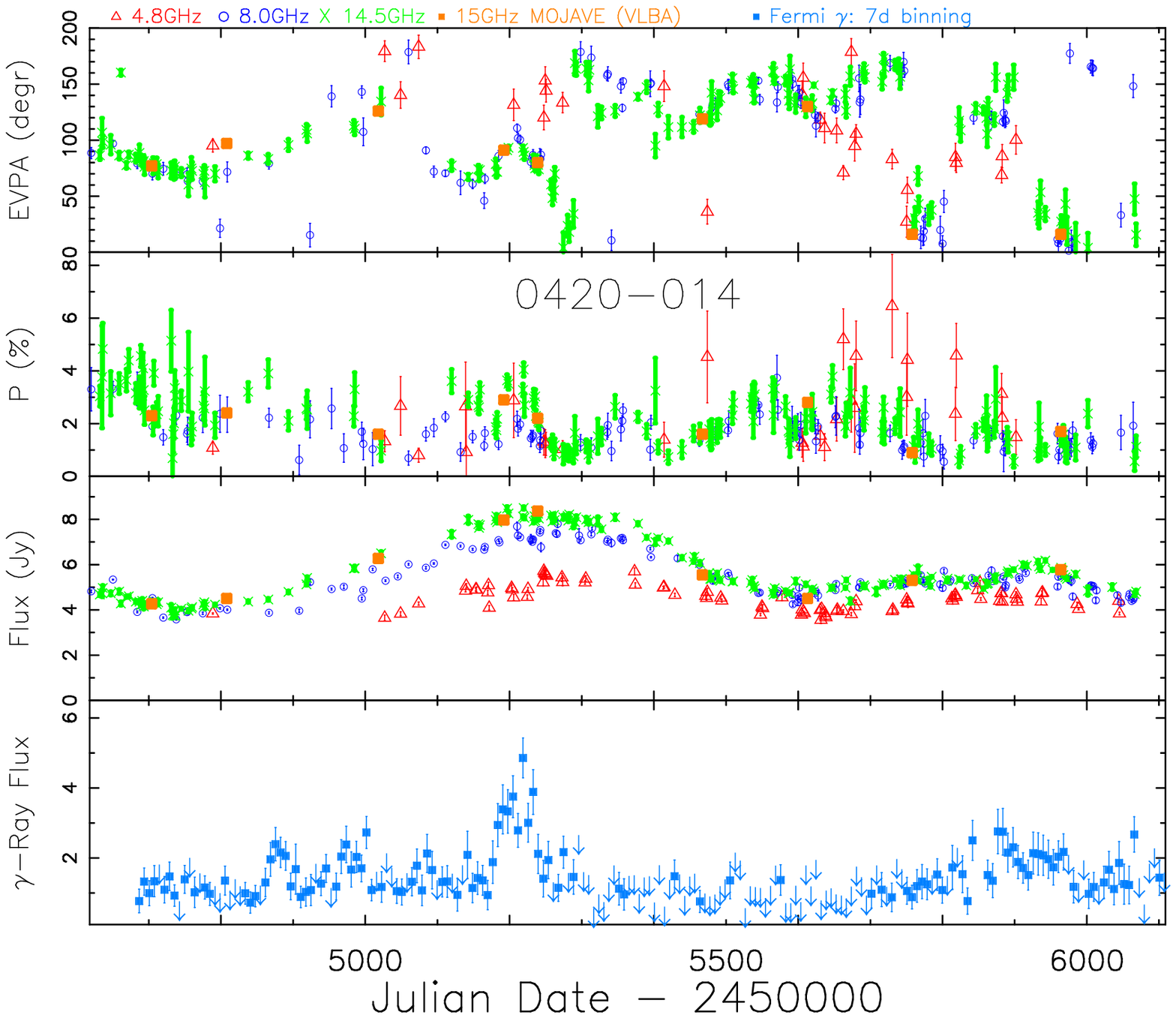}
\caption{Left: Two-week averages of the long-term centimeter-band total flux density, and linear polarization light curves for 0420-014. From bottom to top the panels show total flux density, fractional linear polarization and EVPA at the three frequencies. The frequencies are symbol coded as indicated at the top left. The error bars are 1$\sigma$ error estimates. The  EVPAs (top panel) are restricted to a range of 180$^{\circ}$ due to ambiguities in the reduction procedure; they have exhibited a series of sharp, systematic changes lasting from several months to years.  Right: A blowup showing daily-averaged UMRAO data (panels 2-4) including the time period since the launch of {\it Fermi} (2008.4 through 2012.5). In the linear polarization plots, following the convention adopted in \citet{all85b} data are included only if the associated error for the EVPA determination is less than 14$^{\circ}$; this corresponds approximately to a 2$\sigma$ measurement. The lower panel shows the weekly-binned $\gamma$-ray photon flux in units of photons/s/cm$^{2}\times$10$^{-7}$. The upper three panels include the source-integrated MOJAVE measurements for comparison.   Color versions of the plots are available in the on-line version of the journal for all figures.
\label{fig-1}}
\end{center}
\end{figure}

The long-term centimeter-band total flux density and linear polarization data obtained by the UMRAO monitoring program are shown in the left plot of  Figure~\ref{fig-1} for the LSP QSO 0420-014 (PKS 0420-01; 2FGL J0423.2-0120) at redshift z=0.9161. A precessing binary black hole model has been proposed for this object based on an analysis of 9 epochs of archival VLBI data  at 3.6 centimeters during the time period 1989 - 1992 \citep{bri00}; this work includes the UMRAO total flux density variability data from 1982 to 1994.  Note that nearly-continuous, high-amplitude variability has persisted over decades in both total flux density and in fractional linear polarization. A peak amplitude of S$\approx$14 Jy at 14.5 GHz was attained in late 2003 making this source one of the brightest blazars historically in the centimeter band. Temporally-resolved outbursts in fractional linear polarization occurred throughout the monitoring period, reaching a peak value of 5\%. This low fractional linear polarization compared with the value of $\sim$75\% expected for a canonical synchrotron source is characteristic of the sources in the UMRAO program  and has been used to argue for the presence of a turbulent magnetic field during a quiescent, pre-flare stage. While a source-integrated rotation measure (hereafter RM) of -13 rad m$^{-2}$ \citep{rud83} determined from VLA observations has been included in the linear polarization data plotted, this small value produces a negligible shift in the observed EVPA from the intrinsic value. To test whether inclusion of VLBA-determined  values would reduce the spread in the EVPAs in this source and in the other two sources, we generated light curves incorporating a range of rotation measure values taken from the literature, e.g. \citet{hov12}. The EVPA-frequency offset was smallest for 0420-014 and OJ~287 using the VLA-determined values from \citet{rud83} and when assuming no correction in the case of 1156+295 for which no VLA values are available from either \cite{rud83} or \citet{rus88}. The exercise confirmed that significant RM corrections due to external Faraday Rotation are not required for the data set presented here.

Complementary, long-term spatially-resolved VLBI imaging data from both the MOJAVE 15.4 GHz  and the 43 GHz Boston University program are available publicly. The MOJAVE data for this source commence in mid-1995, with a typical cadence of a few observations per year; these imaging data track the motion of 4 components from 1995 through 2011; based on these data a maximum apparent component speed of $\beta_{max}$=5.74c has been obtained \citep{lis13}. A  parsec-scale jet orientation of 183$^{\circ}$  from MOJAVE measurements is presented in \citet{kar10} which aligns with a large-scale VLA position angle of 183$^{\circ}$. However, at least at 43 GHz, the projected parsec-scale jet orientation varies with time. The data from the Boston University program identify a significant change in direction from  280$^{\circ}$ in 1998-2001 to 100$^{\circ}$ degrees from 2008 to the present time \citep{tro13}. These 43 GHz data which resolve the inner jet region identify that 8 knots have been ejected since early 2009. The derived values of $\beta_{app}$, the component speed, are considerably faster than those identified from the MOJAVE images, but these components lie within the unresolved core region at 15 GHz.  However, the 15 GHz data are able to follow the motions of the individual components to greater distances downstream in the jet.
  
In the $\gamma$-ray band the source was detected by EGRET in the 1990s, with high significance (test statistic TS=46) during the viewing period centered on late February, 1992 \citep{har99} (1992.16) which coincides with a plateau in a centimeter-band flare. A maximum flux (E$>$100 MeV) of 0.45 $\times$ 10$^{-6}$ cm$^{-2}$sec$^{-1}$ is reported in  \citet{mon95}. Shortly after the launch of $\it Fermi$ the source flared in the radio-band.
The event modeled as part of our program is shown in the right plot of Figure~\ref{fig-1}.   For bins with TS$\leq$ 10, a 2-$\sigma$ upper limit is shown.  Several low-level $\gamma$-ray flares were detected prior to the main flare which started near JD 2455170 (2009.93). Panels 2-4 show daily averages of the UMRAO total flux density and linear polarization monitoring observations at 14.5, 8.0, and 4.8 GHz. Substructure is clearly apparent in the fractional linear polarization light curve indicative of emission from individual flares which are blended in the source-integrated, single-dish, total flux density observations. The centimeter-band linear polarization during the $\gamma$-ray flaring exhibits  monotonic changes in the EVPAs through tens of degrees, and associated increases in the fractional LP; these are consistent with the signature of the passage of a shock through the emitting region.

\subsection{Long-term Variability in OJ~287}

\begin{figure}
\begin{center}
\includegraphics[scale=0.50]{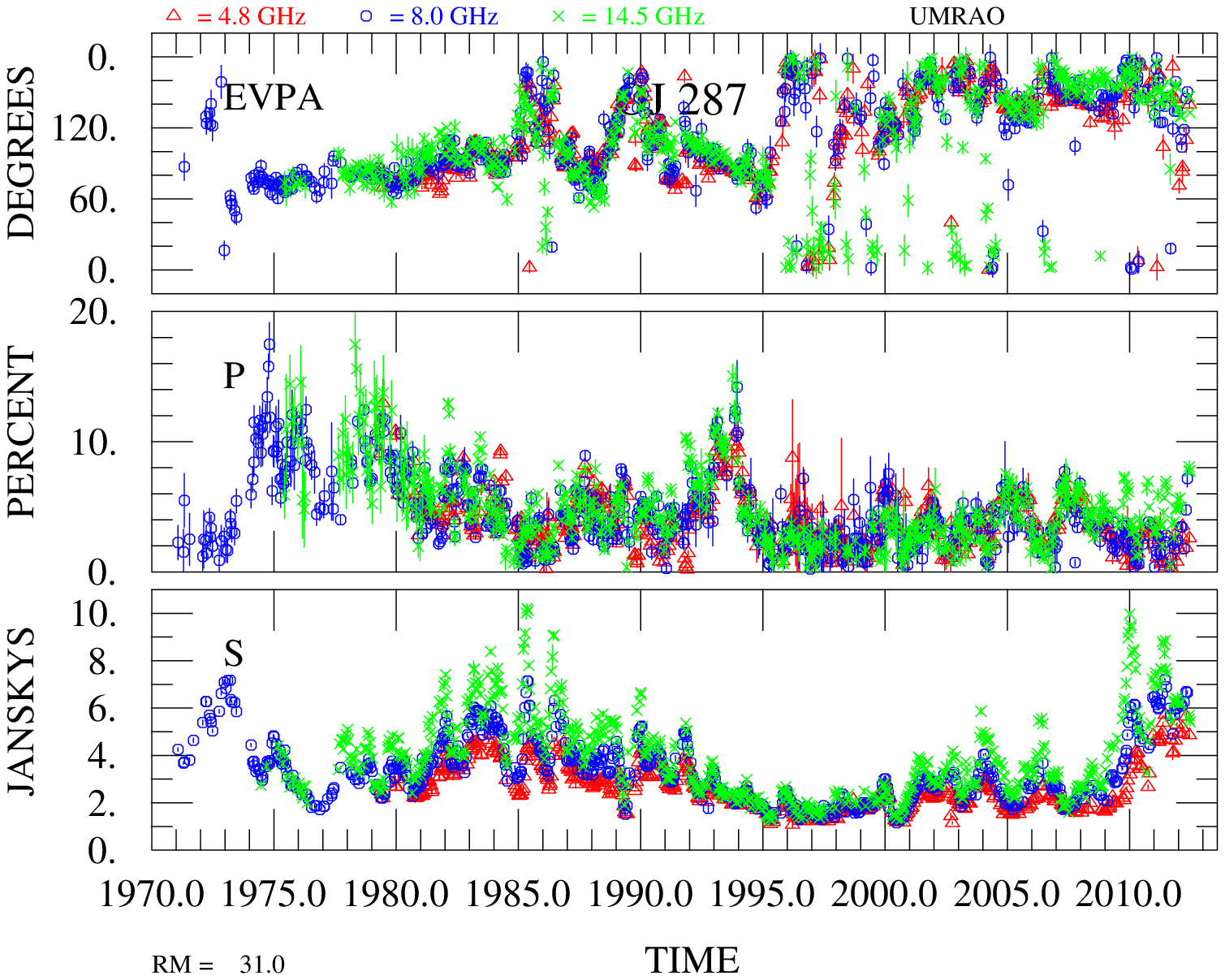}
\includegraphics[scale=0.40]{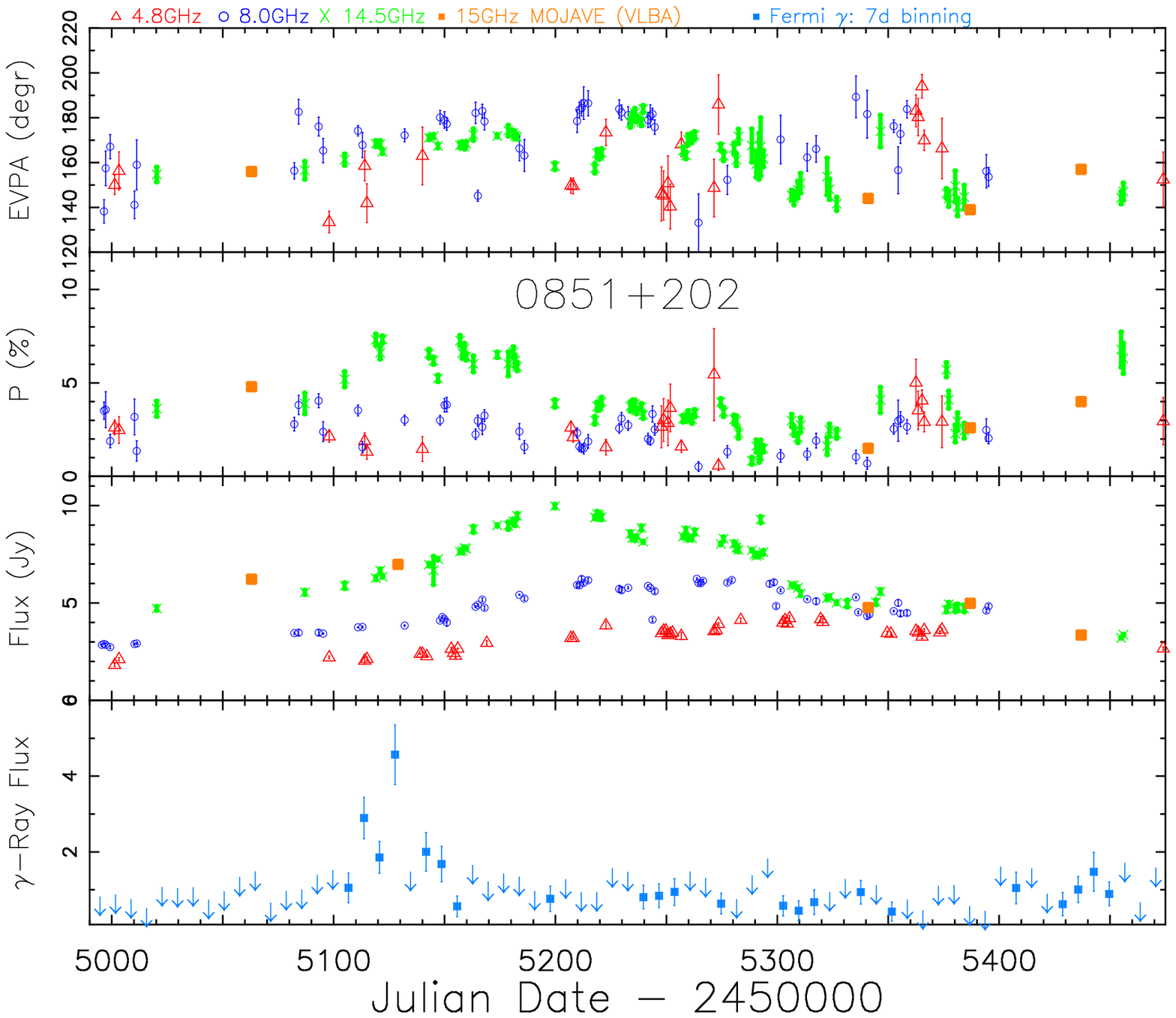}
\caption{Left: Two-week averages of the long-term centimeter-band total flux density and linear polarization light curves for OJ~287. The symbols are denoted as in Figure 1. A source-integrated rotation measure of +31 rad m$^{-2}$ \citep{rud83} has been applied to the linear polarization observations. Right: A blow-up of the time period which includes the event modeled. The format is as in the previous figure. \label{fig-2}}
\end{center}
\end{figure}

Two-week averages of the UMRAO total flux density and linear polarization data are shown in Figure~\ref{fig-2} (left) for the ISP BL~Lac object OJ~287 (0851+202; 2FGLJ 0854.8+2005)  at z=0.306. OJ~287 has been identified as a potential binary black hole based on historical optical data extending back in time to the 1890s; both a 12-year period attributed to the perturbation of the accretion disc of the primary by the secondary, and a 60-year period identified with a precession cycle have been proposed, e.g \citet{val12}. Cyclic variations in the fractional linear polarization in the optical band with a time-scale of 2 days are discussed in \citet{dar09}. 
A wavelet analysis of more than two decades of the UMRAO total flux density and linear polarization data (1971 -- 1998) identified two persistent signals -- a longer time scale period of $\approx$ 1.66 years associated with the quiescent jet, and a shorter periodicity of $\approx$1.2 years associated with the series of large flares which occurred during the 1980s \citep{hug98}. The fractional linear polarization exhibits well-defined nulls between flares. A new cycle of intense activity commenced circa 2000 following a relatively quiescent period. While the maximum amplitude of fractional linear polarization is only 5\%  at 14.5 GHz for 0420-014, similar to values typically found for UMRAO program sources, for OJ~287 we detected unusually high
values of fractional linear polarization which are in the range P$\approx$15-18\% at 14.5 GHz; these high levels were reached during relatively quiescent total flux density levels and are not associated with large flares. One of these peaks occurred in fall 1993 during the operation of EGRET; at this time the total flux density was only 2 Jy. 

The MOJAVE data at 15.4 GHz identify 5 primary moving components and a maximum apparent component speed, $\beta_{max}$, of 15.14c based on seven moving features\footnote{http://www.physics.purdue.edu/astro/MOJAVE/velocitytable.html}; from an analysis of the variation of the projected innermost jet position angle on the sky with time, a monotonic swing in jet position angle of a few degrees per year from 1995 to the end of 2010 is discussed in \citet{lis13}. Campaign VLBA data at 43 GHz obtained in October -- November 2005 and March -- April 2006 have been modeled with a fast spine and a slow sheath \citep{dar09} with a  maximum Lorentz factor of 16.5 for the spine, and a Lorentz factor of only 5 for the sheath. The presence of non-periodic jet wobble has recently been suggested based on longer-term 43 GHz VLBA  data \citep{agu12}. Differences in the behavior at 43 and 15.4 GHz are attributed to the different angular resolutions of the data sets \citep{lis13}. As discussed in a later section of the paper, there is also significant opacity in the source in the centimeter-band emission region which may contribute to these differences.

In the $\gamma$-ray band, this source was detected by EGRET with 3$\sigma$ detections (TS=9) during CGRO pointings centered on late September 1992 and in mid-November 1994 \citep{har99}, but no strong EGRET detections were reported during the epoch of the high fractional linear polarization in 1993 apparent in the UMRAO data. During the operation of EGRET the radio-band total flux density was on the decline. In contrast,  the source has been unusually bright in the past few years, reaching an amplitude of S=10 Jy, which is just below the highest amplitude measured historically by the UMRAO program data. Only one strong $\gamma$-ray flaring period (photon flux $>$5 $\times$ 10$^{-7}$ photons s$^{-1}$cm$^{-2}$) was detected by {\it Fermi} during the time window shown in Figure~\ref{fig-2} (right).

\subsection{Long-term Variability in 1156+295} 

\begin{figure}
\begin{center}
\includegraphics[scale=0.48]{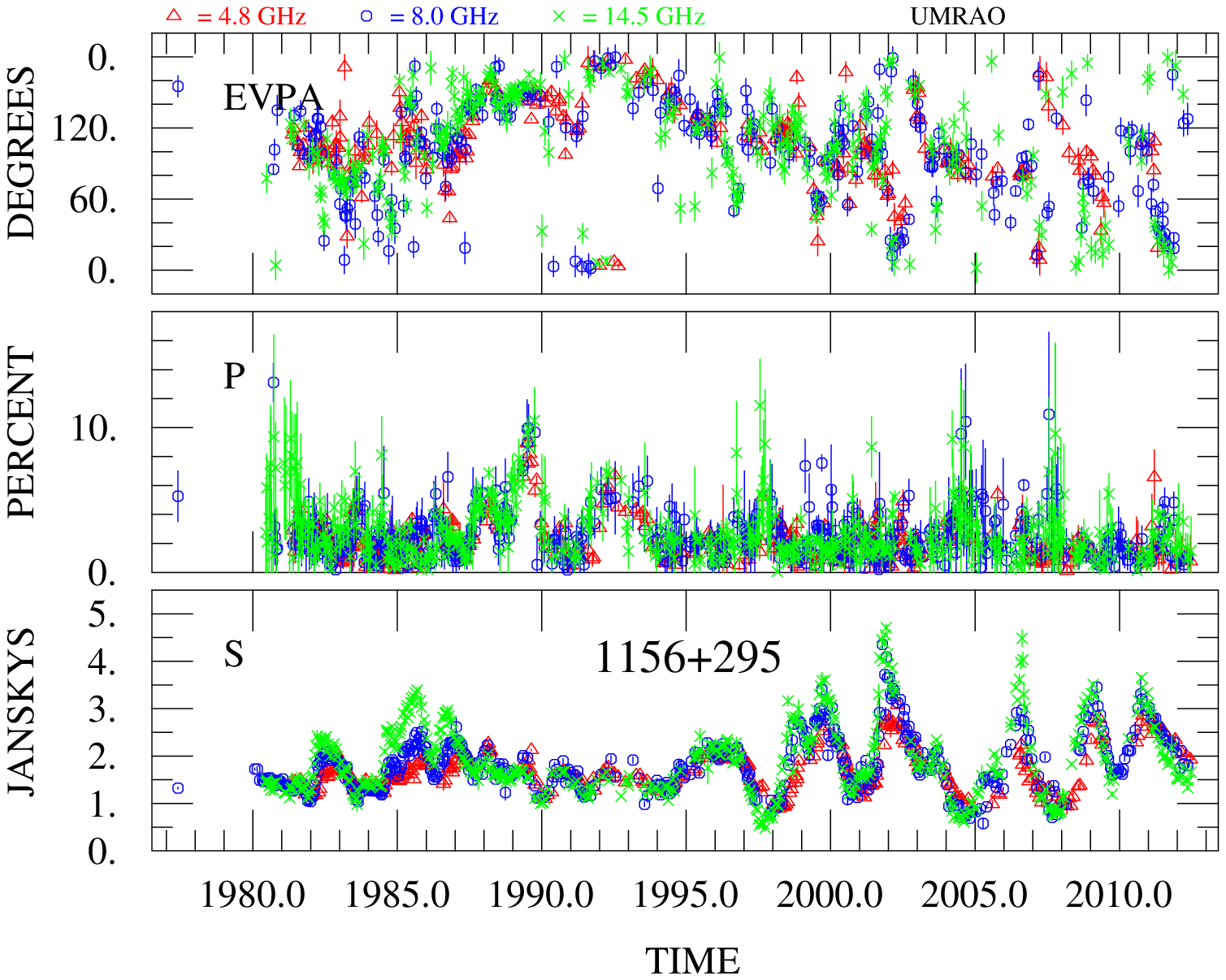}
\includegraphics[scale=0.42]{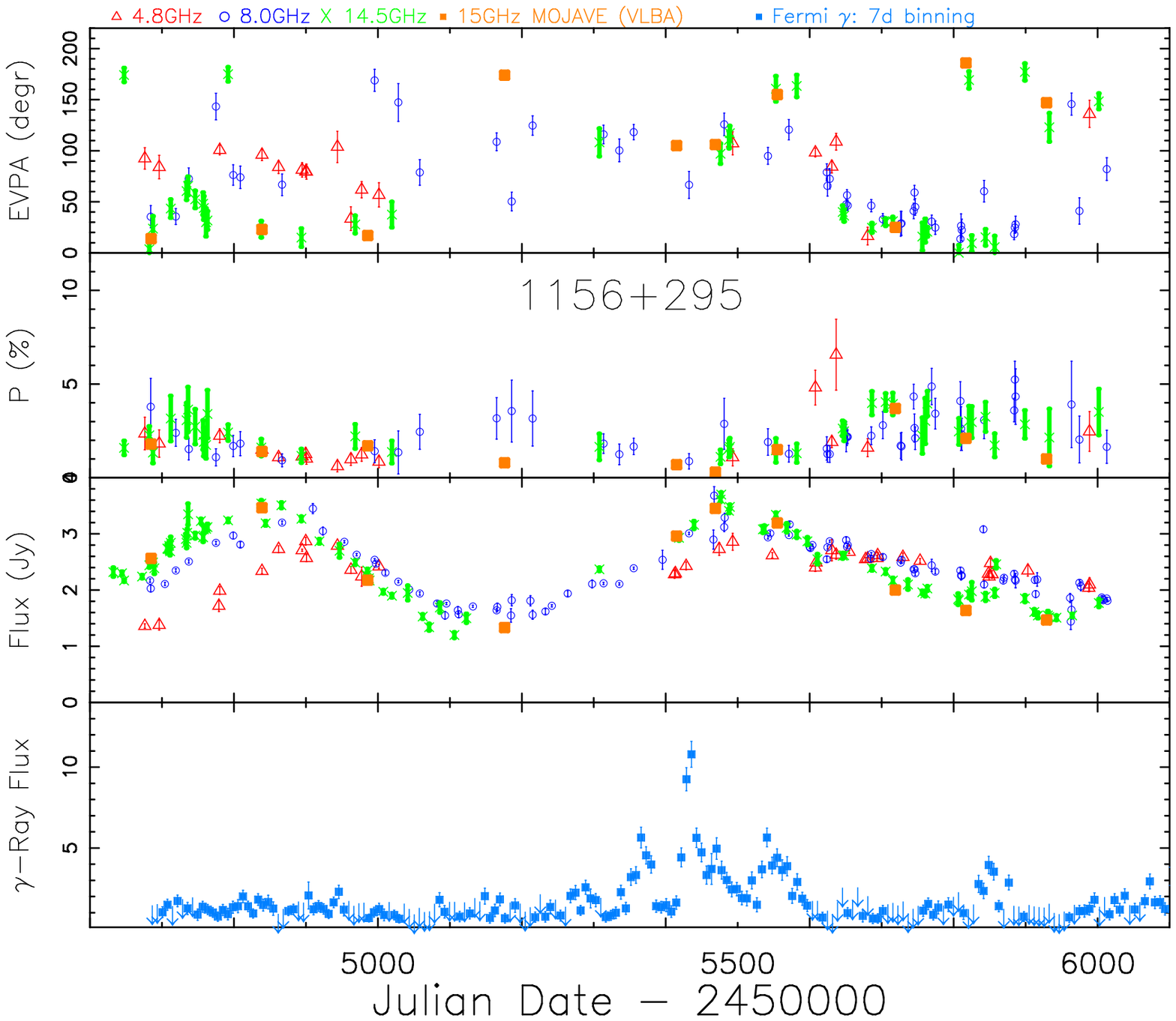}
\caption{Left: Two-week averages of the long-term UMRAO data for 1156+295. A source-integrated VLA rotation measure is not available from \citet{rud83}. No frequency-dependent $\lambda^2$ separation of the EVPAs indicative of Faraday Rotation is apparent in the UMRAO data. Right: A blow-up of the data during  2008.4 - 2012.5. An analysis of the radio-band flare to the right with a temporally-associated $\gamma$-ray flare is modeled in this paper. \label{fig-3}}
\end{center}
\end{figure}

The LSP, flat spectrum radio quasar 1156+295 (TON 599; 4C +29.45; 2FGL~1159.5+2914) at z=0.725  has exhibited nearly-continuous, centimeter-band flaring over the $>$30 year time period shown in Figure~\ref{fig-3} (left). Until 1998.0 the amplitude of the total flux density variations was relatively modest. Subsequently the variability changed in character, and several well-separated flares in total flux density occurred. During these events the total flux density is characterized by a flat spectrum, and the events exhibit an unusual shape characterized by nearly equal rise and fall times, reminiscent of strong, centimeter-band  events in 1510-089 \citep{mar03}. The time segment since the launch of $\it Fermi$ includes two of these large and distinctive flares. Overall, the fractional linear polarization has exhibited  a long-term base level near 2\% and reached a maximum fractional linear polarization of about 10\%. The EVPAs show no preferred orientation, and they exhibit a large range of values in the light curves which show complex structure. 

A maximum apparent component speed of $\beta_{max}$=24.59c has been found from MOJAVE data, making the flow in this source the  fastest of the three sources studied here, assuming that this pattern speed is an indicator of the bulk flow speed.
Bending of the relativistic jet on VLBI and VLA scales away from the observer was invoked to explain earlier X-ray variability \citep{McH90}. 

In the $\gamma$-ray band, EGRET obtained a highly-significant detection with TS$\geq$36 during early January 1993, and an additional detection with TS$>$36 with lower photon flux was obtained in late April--early May 1995 \citep{har99}. The $\it Fermi$ time window shown in the lower panel of the right plot in Figure 3 captures two similar radio-band flares. The {\it Fermi} data train commences just after the start of the first outburst, and there is no indication of an enhanced flux density level in the early measurements indicative of a prior flare. The first radio-band flare does not have a clearly-associated $\gamma$-ray counterpart (an example of an orphan flare), while the second, slightly flatter-spectrum event does. While changes in the flow may have occurred between the times of these radio-band events upstream of the centimeter-band region, no change in opacity within the centimeter-band emission region is indicated by the UMRAO data.  Further, only modest changes in the projected inner jet position angle between 2008 and 2010 are identified by the MOJAVE data \citep{lis13}, and hence geometrical differences do not appear to account for the occurrence of the radio-band flare with no $\gamma$-ray counterpart.

\section{Radiative Transfer Modeling}

\subsection {Model Framework and Assumptions}

The formulation and assumptions adopted in the radiative transfer models incorporating a propagating shock scenario and used in the analysis presented here are described in detail in  \citet{hug11}. We summarize some of the salient features here.  

The models assume that a power law distribution of radiating particles permeates the quiescent flow and that the density distribution of these particles follows a power law of the form 
$n(\gamma$)d$\gamma$ = $n_{0}\gamma^{-\delta}$d$\gamma$ where  $\gamma$ $>$ $\gamma_{i}$.
The low-energy cutoff of this distribution is specified in terms of a thermal\footnote{pertaining to random motions, not the bulk flow} Lorentz Factor (LF), $\gamma_{i}$, and a fiducial value, $\gamma_{c}$, indicates from where in the particle spectrum the observed emission arises. The power law index, $\delta$, ($\delta$=2$\alpha_{thin}$+1; $\alpha_{thin}$ is the optically thin spectral index) is fixed, and the density constant, $n_0$, is assumed to fall off due to the adiabatic expansion of the flow.

The models assume the presence of a passive, turbulent magnetic field in the synchrotron-emitting region of the jet before the passage of the shock. The magnetic field structure in this region is represented by cells with randomly-ordered field directions. 
The passage of a shock through this region produces a compression which increases the particle density and the magnetic field energy density (and hence the emissivity). The degree  of order of the magnetic field is increased by the compression associated with the passage of the shock, and this leads to an increase in the fractional linear polarization. 

The formulation of the models allows for shocks to be oriented at any angle to the flow direction \citep{hug11} and builds on earlier work  restricted to the  special case of transverse shocks \citep{hug85,hug91}. Each shock's orientation is specified by two angles; these are its obliquity, $\eta$, measured with respect to the upstream flow direction, and the azimuthal direction of the shock normal, $\psi$.  Earlier work showed that the simulations are relatively insensitive to changes in the azimuthal direction, however. The shocked flow is characterized by a compression factor, ($\kappa$),  a length (l) defined as the evolved extent of the shocked flow, and the shock sense (forward or reverse: F or R).  The outburst seen in the variability data is associated with the propagation of the region bounded by the limits of the shocked flow. To simplify the computations, each shock is assumed to span the entire cross section of the flow, and it is assumed to propagate at a constant speed.  Additionally, multiple shocks contributing to a single outburst are assumed to have the same orientation, and subsequent shocks contributing to an outburst do not shock pre-shocked plasma. Inclusion of retarded-time effects for test cases revealed that only small differences in the simulated flare shapes and spectral behavior result {\citep{all13a}, and hence this effect is neglected in the simulations presented here.

To reproduce a well-defined EVPA in the quiescent state a small fraction of the magnetic energy (2\%) is initially assumed to be in an ordered component of the magnetic field; support for this assumption comes from the low levels of fractional linear polarization identified during relatively quiescent states in the UMRAO data \citep{hug11}. An important factor in the modeling is the observer's viewing angle. Initial estimates for this parameter for each source were taken from VLBI studies (Table~\ref{tbl-1}) limiting our need to explore parameter space. Table~\ref{tbl-1} additionally includes the maximum apparent component speed, $\beta_{max}$, determined from 15.4 GHz MOJAVE data, \citep{lis13} in column 2; the  variability Doppler factor, $\delta_{var}$ in column 3, derived from fits to  archival Mets\"ahovi single-dish monitoring data at 22 and 37 GHz \citep{hov09}; a flow Lorentz factor, $\Gamma$, and viewing angle, $\theta_{var}$,  in columns 4 and 5 respectively computed from these data following the prescription in \citet{hov09} and using the revised apparent speed determined from longer-term MOJAVE data; and, in columns 6 through 8 the Doppler factor, Lorentz factor, and viewing angle determined by \citet{jor05} from 43 GHz VLBI data which probe the flow upstream of the centimeter-band emission region. These results provide useful comparisons with our derived parameters.

\subsection{Modeling Procedure}
A complication in analyzing single-dish, centimeter-band light curves is the blending of emission contributions from individual, evolving flares. In general the UMRAO light curves trace out an envelope over activity which may include contributions from several flares/shocks. Past work on separating blended flares, particularly in the analysis of mm-wave data, has often assumed a generic flare shape with an exponential decay \citep{val99,leo11}. This exponential shape, however is not a good match to the centimeter-band events, and, further, the method requires knowledge of the baseline flux level for removal of this emission contribution prior to the fitting process. Instead, we have resolved the blended events into individual flares by using a combination of the structure apparent in the total flux density and linear polarization light curves at our highest frequency, 14.5 GHz, where they are best resolved, and the expected flare shape for a single shock based on a library of simulated light curves generated as part of this work. Example simulated light curves for a single shock are shown in \citet{hug11}.

The number of shocks contributing to a modeled event, established by this
deconvolution procedure, is fixed during subsequent modeling. A quiescent
flow Lorentz factor and shock sense (always `forward' for the sources
discussed in this paper for the reasons discussed in  \S5) are chosen based on the expected viewing angle
and the observed proper motions as discussed in \S 3.1. The component proper
motions predicted by the model depend on the quiescent flow Lorentz factor
and shock strength, so a typical shock compression is adopted in setting
up the initial state of the model. An initial shock obliquity is chosen by
inspection of the range of EVPA  change displayed by the data, using the results
of \citet{hug11} to estimate the obliquity needed to produce a match to the data.

The fiducial Lorentz factor ($\gamma_c$) is arbitrarily set to $1000$, and
the cutoff Lorentz factor ($\gamma_i$) is set to a high value (typically $50$
or $100$), to ensure negligible internal Faraday effects. The optically-thin
(frequency) spectral index, $\alpha$, is fixed at a value of 0.25 based on the rather
flat total flux density spectra exhibited by most UMRAO sources even when there is no evidence
that opacity is important.

The onset time, length, and strength (compression) of the shocks are then
adjusted iteratively, in an attempt to fit the event shape in total flux
(quantitatively, the flux increase at peak, the amplitude of the change in the total flux density $\Delta S$,
the spectral shape
when the emission is most opaque, and the amplitude and position of structure within
the light curve) and the observed fractional linear polarization (quantitatively,
the peak value during the event). If a satisfactory match to the data
cannot be achieved, the viewing angle is adjusted, and the process repeated.
The quiescent flow Lorentz factor is adjusted similarly if no viewing
angle is found which yields a good match to the data. For a given quiescent flow Lorentz
factor the model fractional linear polarization is very sensitive to the viewing angle,
so that a change in the viewing angle can be used to refine the model fractional linear
polarization while leaving the total flux light curves largely unchanged;
importantly, the value of the viewing angle is very well-constrained by the modeling.

Structure in the observed fractional linear polarization light curves, and trends
in the EVPA are then examined by eye (in general the observed EVPA light curves display very
complex behavior), and the shock obliquity and the low-energy spectral cutoff
are adjusted to reproduce these features. A change in the
cutoff (introducing significant internal Faraday effects) can modify the
model fractional linear polarization, requiring further iteration of the shock
parameters, viewing angle, and the quiescent flow Lorentz factor.

We emphasize that the intent of the fitting procedure is to ascertain whether the model can
reproduce the general features of the data. We do not provide
proof of uniqueness of the models; nor can we quantify a goodness of fit,
as the final phase of modeling attempts to refine the model parameters
using trends in the EVPA variability established by inspecting the data.

\section {Determination of Source Parameters from the Modeling}

We have used the occurrence of a bright $\gamma$-ray flare to identify the time windows selected for the modeling. Our underlying premise is that the same disturbance is responsible for the flares in both bands, and that by modeling the radio flares, we can set constraints on the physical conditions at or near to the $\gamma$-ray emission site.  We have not carried out cross correlations to establish causality between the radio-band outbursts modeled and the temporally-associated $\gamma$-ray flares because a large body of past work has shown that there is a  strong dependence of the correlation significance on the time window selected when data trains of only a few years are used \citep{max12,max13,ric13}. Additionally the time scales for the variability in the radio and $\gamma$-ray bands are characteristically different (typically hours to weeks at high energies and from months to years in the radio band \citep{all10} hampering the unambiguous association of specific flares even when delays due to opacity are known and included. Instead we adopt the view a priori that the selected flaring events are produced in the same region of the jet flow, and that it is plausible that the cross-band activity during these flares is causally related. This view is based on the observational evidence that: 1) the $\gamma$-ray flares selected are well above the detection level and both bright and persistent in the weekly-binned photon flux data; 2) these $\gamma$-ray flares are associated with unusually strong flaring in the radio-band based on historical centimeter-band measurements obtained over decades, e.g. \citet{val99}; 3) $\gamma$-ray flaring occurs at or near to the rise portion of the radio-band event, consistent with the expectation that the radio-band and $\gamma$-ray events are produced by the same disturbance, e.g. \citet{lah03,leo11}; and 4) there is mounting statistical evidence from VLBI monitoring at 43 GHz of $\gamma$-ray flaring sources that component ejections from the 43 GHz core region are associated temporally with  $\gamma$-ray flares in a statistically significant number of events \citep{jor12} showing a correspondence between $\gamma$-ray and radio-band activity.

\subsection{0420-014}

 The UMRAO data for the event modeled are shown in Figure~\ref{fig-4} (left) which spans the 2-year period 2009.0 to 2011.0. (circa JD 2454832 -- 2455562). These data exhibit the characteristic behavior expected for shocks propagating through the emitting region: an outburst in total flux density, time-associated outbursts in polarized flux, and ordered changes in the EVPA light curve. A characteristic behavior in the total flux density light curves is blending of the contributions from individual flares; these individual flares are better-resolved  in the linearly-polarization light curves as is illustrated by the structure in this fractional polarization light curve. There is a 180$^{\circ}$ ambiguity in each EVPA determination which can affect the interpretation of the temporal changes exhibited by the EVPA data. In general the choice of whether to add or subtract a multiple of 180$^{\circ}$ when generating UMRAO EVPA light curves is based on an algorithm which makes this decision by requiring the smallest point-to-point jump in the time sequence of the EVPAs using the data at all three frequencies, and additionally requiring that differences between consecutive data points be less than 90$^{\circ}$. Further, the range of EVPA values is restricted to 180$^{\circ}$. However, for ease in comparison of the data with the simulated light curves for this source, we relaxed these restrictions on the EVPA range and added multiples of 180$^{\circ}$ to reproduce the character of the simulated light curve shown in Figure~\ref{fig-4} right. Following \citet{all85b} we have only included data in this plot, and in those for the other two sources studied, for which the differential errors in EVPA derived from the standard errors of the Stokes parameters are less than 14$^{\circ}$ (P/$\sigma_P>$2). Several observations were rejected at 4.8 GHz because they did not meet this acceptance criterion: the observations at this frequency were often made during degraded weather conditions, while the best weather was generally reserved for observations at 14.5 GHz. As a consequence, the observed 4.8 GHz  data is more poorly sampled than at the two higher frequencies. The EVPAs during the outburst show monotonic swings in EVPA through $\sim$ 90$^{\circ}$ as expected for the passage of a transverse shock through the emission region. The best-sampled case is the systematic change in EVPA from late 2009 through early 2010. These changes occur during the rise portion or plateau of the individual flares within the outburst envelope, as expected for the shock scenario. While evidence for large, systematic rotations of the EVPA have been found for a few sources based on the UMRAO data \citep{led79,all81}, these rotations have occurred when the total flux density had reached a maximum during an outburst or was declining  and are characteristically different from the linear polarization changes produced by the passage of a shock. Shocks produce an increased degree of order of the initially-turbulent magnetic field and an increase in the fractional linear polarization apparent in the light curves as a polarization outburst. Polarization rotations occur when the fractional linear polarization is relatively small ($<$2\%; they can be produced via a random walk resulting from the evolution of the turbulent magnetic field within the emission region \citep{jon85}. 

Based on the structure in the total flux density and linear polarization light curves, combined with the burst profile shape expected for a single shock, we introduced 3 shocks oriented transversely to the flow direction into the simulations.  The onset times of the shocks included in the simulation are indicated by  upward arrows along the abscissa of the lower panel, while the $\gamma$-ray peak photon flux is marked by a downward arrow at the top of this panel. The parameters specifying each shock start time, the length as a percentage of the flow, the compression factor, and the location of the peak flux in dimensionless time units are summarized in Table~\ref{tbl-2}.  The first shock is weak, the second shock which starts at 2009.6 is the main shock, and the third shock which commences around 2010.0 is the shock which coincides temporally with the brightest $\gamma$-ray flare. Figure~\ref{fig-4} right shows the simulated light curve.  The amplitude of the fluxes   presented here and in the other simulations shown are scaled to match the peak total flux density in the observed light curve. Twenty time steps were used in all of the simulations presented in order to resolve the structure in the light curves during the event modeled.

Comparison of the data with the simulations reveals that the simulations are able to reproduce the global features of the light curves; these include the range in the total flux density at the highest frequency (14.5 GHz), the evolution in the spectral index  of the total flux density during the outburst based on the amplitudes of the total flux density at 14.5 and 4.8 GHz, the amplitude of the change (maximum to minimum) of the fractional linear polarization, and the change in the spectrum of the EVPAs during the evolution of the outburst; the latter is complex. The flow properties adopted in the simulation shown are: an optically-thin spectral index $\alpha_{thin}$=0.25 (fixed for all three sources), a fiducial ``thermal'' Lorentz factor (associated with the random motions of the emitting particles) $\gamma_c$=1000, a low energy cutoff ``thermal'' Lorentz Factor of $\gamma_{i}$=50, a bulk Lorentz factor of the flow $\gamma_{f}$=5.0, and an observer's viewing angle of $\theta_{obs}$=4$^{\circ}$. From the modeling, the derived Lorentz factor of the shocked flow is 8. At a viewing angle of 4$^{\circ}$ the shock transition Lorentz factor (which is the emission feature which one would see moving in the VLBA images) implies
 $\beta_{app}\sim$11c. While the overall spectral behavior is very well reproduced by the simulation, note that there are some differences, including the range of the EVPA swing. The extent of the swing is best judged from the behavior while the trailing shock (shock 3) dominates. In the data the range of the swing is through approximate 135$^{\circ}$. A comparable  EVPA swing occurred during the first shock; however, with the sampling of the data, the time at which the shock enters the flow and becomes manifest is not as well defined in the light curve. Additionally, the null in the observed fractional linear polarization light curve and an associated sharp variation in the EVPA light curve near 2010.6 are not reproduced by the simulation. These  differences between the data and model may result from the adopted simplifying assumptions in the model formulation.

This source is included in both the MOJAVE and the 43 GHz BU VLBA programs, and it is instructive to compare the shock start times with those of the propagating jet components apparent in these VLBI images. A detailed  analysis of the 43 GHz imaging data during the $\gamma$-ray flaring is presented in \citet{tro13} based on monthly-sampled data.  A comparison between onset times and apparent speeds of the shocks and components is presented in Table~\ref{tbl-3}. Recall that the shock onset times correspond to the time when the leading edge of the shock enters the flow; this is not the same parameter as the time at which the separation of the component from the radio-band `core' occurs. At 43 GHz, five  components have been ejected during the time period 2009.0 through 2011.5, and 4 of these occur during the time window included in the simulation. Within the error bars, the injection times for  components K2-K4 are consistent with the onset of our three shocks. The 4th component is very fast, and there is no corresponding shock required from our analysis. Overall, however, this comparison reveals an excellent correspondence between the moving emission features and the model shocks. Note that the model $\beta_{app}$ and the observed component speed are nearly equal for shocks 1 and 3.

\subsection{OJ~287}

The radio-band event which was selected for modeling occurred during the 10-month time period from September 2009 to July 2010 and commenced just after the source exited from the summer 2009 sun gap (a time at which it was located too close to the sun for observation by UMRAO). The observed light curve is shown in the left plot in Figure~\ref{fig-5}. A large, self-absorbed total flux density flare took place during this time window which temporally coincides with a large $\gamma$-ray flare as shown in Figure~\ref{fig-2} (right). A large flare in fractional linear polarization is also apparent in the UMRAO monitoring data which reached a maximum fractional linear polarization near 8\%  during the rise portion of the outburst. The EVPAs show systematic changes, most apparent at 14.5 GHz; these changes are through a smaller range of EVPA values than observed for 0420-014, and the pattern is consistent with the presence of oblique rather than transverse shocks.  The simulation shown in  Figure~\ref{fig-5} (right) incorporates 3 forward-moving shocks; the parameters specifying these shocks are tabulated in Table~\ref{tbl-4}. Comparison of the model light curves and the data shows that temporally-resolved swings in EVPA are associated with the entry of the shocks into the flow.  As in the case of 0420-014, these EVPA swings occurred during the rise or plateau phases of the blended, total flux density light curve. Note that the EVPA swings at 14.5 GHz in the simulation and in the data are through a comparable range of about 30$^{\circ}$. The range of the swing at 4.8 GHz, however, is through a larger range in both the data and in the simulation.  There is an indication of a swing through a similar range at 8 GHz for the observed changes associated with shocks 1 and 2, but unfortunately, the data are not sufficiently well-sampled at either 4.8 or 8.0 GHz to capture the full range of the observed swing. Additionally, the presence of larger swings at 4.8 and 8.0 GHz is masked by the lower signal-to-noise of the data. The larger swings in the simulated light curves at the two lower frequencies through approximately 90$^{\circ}$  are the result of Faraday effects; at 14.5 GHz these effects are reduced, and the EVPA swing reflects the underlying oblique shock. While the modeling identifies Faraday effects within the source during the time period modeled, note that the separation of the EVPAs in the data plot does not follow a simple $\lambda^2$ relation. We attribute the observed spectral evolution of the EVPAs to the complex effects of shocks within the flow, Faraday effects, and opacity. In this source there is evidence for substantial self-absorption in the total flux density light curve, and this important feature of the total flux density is well-reproduced by the simulated light curve shown in the right plot. This model additionally reproduces the range of change in total flux density (2-9 Jy), the spectral evolution of the total flux density (inverted near outburst peak, flattening in the decline portion of the event), the approximate time delay of the peak flux at each frequency; the range of the change in fractional linear polarization (0-6\%); and  a swing of the EVPAs through a smaller range of values than expected for a transverse shock. In the simulation shown we adopted a fiducial ``thermal'' Lorentz factor (associated with the random motions of the emitting particles) $\gamma_{c}$=1000, a low energy cutoff Lorentz Factor $\gamma_i$=20, a bulk Lorentz factor of the flow $\gamma_{f}$=5.0, and an observer's viewing angle of $\theta_{obs}$=1.5$^{\circ}$. 

As in the case of 0420-014, it is useful to compare these modeling results with the parameters identified from VLBI imaging studies. The long-term MOJAVE data identify a range of $\beta_{app}$ from 0.438c (a nearly stationary, long-lived component) to 15.13c \citep{lis13}; component 11, ejected nearest to the launch of {\it Fermi}, has a speed of 6.58$\pm$0.56c. The Lorentz factors of the modeled shocks, denoting the speed of the shock transition, are near 20c; these show  relatively little spread since the compressions are similar, at least for shocks 2 and 3.  The derived values of $\beta_{app}$ from the modeling are 17c. We note that there is evidence for a change in the flow orientation with time \citep{lis13} and that the modeled flow seen at angles between 0.5 and 2.5$^{\circ}$ would produce $\beta_{max}$  values of 5c to 20c consistent with the range of MOJAVE component speeds. This result suggests that geometric effects may be important in interpreting the data for this source. As noted earlier, there are differences between the flows at 43 and 15 GHz based on the VLBI monitoring data  \citep{lis13}.

\subsection{1156+295}

In Figure~\ref{fig-6} we compare the data  (left) and the simulation {(right) for the outburst modeled in 1156+295.  The shock parameters are given in Table~\ref{tbl-5}. In this source 4 forward shocks were required. The shock compressions range from 0.5 to 0.8, while the lengths of the shocked flows are identical.  The simulation shown reproduces the relative amplitudes of the total flux density at the three frequencies and the spectral evolution of the total flux density, including the cross-over of the 14.5 GHz total flux density relative to the values at the other two frequencies during the outburst decline, the peak amplitude (6\%) of the fractional linear polarization, and the monotonic increase in the linear polarization during the decline portion of the total flux density outburst.  The simulated light curve shows two swings through approximately 90$^{\circ}$. The first is associated with the entry of shock 1 into the flow.  The data during this portion of the outburst are not sufficiently sampled to define the range of the swing using the 14.5 GHz alone. However, the combined 14.5 and 8 GHz data are consistent with a swing through a comparable range. The second swing terminates as the amplitude of the fractional linear polarization begins a new rise. This characteristic behavior is well matched by the simulation. However the swing is through approximately 120$^{\circ}$ exceeding the theoretical prediction of 90$^{\circ}$. In addition to the proposed explanations for 0420-014, here shock interactions may contribute to the complexity of the flow \citep{all14}.

 To characterize the flow, we adopted a fiducial ``thermal'' Lorentz factor $\gamma_{c}$, the low energy cutoff Lorentz Factor $\gamma_i$=50, a bulk Lorentz factor of the flow $\gamma_{f}$=10.0, and an observer's viewing angle of $\theta_{obs}$=2$^{\circ}$. The bulk Lorentz factor of the flow is the highest of the three sources modeled, a result constant with the values listed in Table~\ref{tbl-1}. This source is included in both the BU and the MOJAVE VLBA monitoring programs. Component speeds are presented for 5 jet components in \cite{lis13}. These range from 1.65$\pm$0.30c to 24.6$\pm$1.9c.  Note that the model $\beta_{app}$ of 22c is in very good agreement with this VLBI result based on contemporaneous data. 

 \subsection{Effect of Changes in the Parameter Values on the Simulated Light Curves: axial B field \& low energy cutoff}
 In Paper II (in preparation) we present a detailed discussion of the effect of changes in the input parameters on the simulated light curves. However, to aid the reader in assessing the sensitivity of the simulations to the choice of these parameters and hence to judge the quality of the fits, we illustrate the effect of independently changing the fraction of energy in an ordered axial component of the magnetic field for all three sources, and the effect of increasing the cutoff Lorentz factor for OJ~287 to match the value of 50 found for the other two sources. 

During the analysis of the third source in the trio presented, 1156+295, we found that an increased contribution from an ordered axial magnetic field component was required to reproduce the features of the observed light curves.  Motivated by this result, an increased ordered component of the magnetic field was included in the simulations for OJ~287 and 0420-014, and these results are presented in this paper. These new light curves are better able to reproduce the structure within the observed light curves and the amplitude of the linear polarization compared with our earlier results in \citet{all13a}. To illustrate the affect of this modification, we show in Figure~\ref{fig-7} (left and right)  and Figure~\ref{fig-8} (left) light curves generated assuming the weakest axial magnetic field able to yield a well-defined EVPA in the quiescent state (2\%).  For 0420-014 comparison of the observed light curves with the simulations based on the adopted model (Figure~\ref{fig-4}) reveals that the fractional linear polarization is 50\% too high at all three frequencies (nearly 6\% at 14.5 GHz compared with the observed value of 4\%), and that there is no significant structure in the fractional linear polarization contrary to the data and the adopted model; additionally the EVPA is constant from early in the rise of the total flux, again contrary to both the data and adopted model for this event. For OJ~287 as shown in Figures~\ref{fig-8} (left) adopting  the minimal mean field yields a  peak fractional linear polarization that is similar in amplitude to the data and to the adopted model; however,
the value remains high until late in the outburst, while in the data and adopted
model it drops by almost a factor of two well before the peak of the total flux.  Again,
the alternate model light curve is smoother with less `structure' than is exhibited by the data. Further, subsequent to the initial rise in total flux density
the EVPA has a smooth variation spanning only tens of degrees, not reproducing
the rapid swings apparent in the observed light curve. In the  case of 1156+295, the simulated fractional linear polarization exceeds 10\% and the time variation in the EVPAs is too flat when compared with the observed features. Hence adopting a value for the  axial magnetic field above the minimum level improves the ability of the simulations to reproduce the detailed structure apparent in the linear polarization light curves in this source.

  For OJ~287 we show in Figure~\ref{fig-8} (right) the simulated light curve produced by increasing the low energy cutoff from 20 to 50 to match the values adopted for the other two sources, with all other parameters as tabulated for the adopted model. Here the peak fractional linear 
polarization is about 25\% too high compared with the data and the adopted model, and it remains high until late in
the outburst, while in the data and adopted model it drops by almost  a factor of two
well before the peak of the total flux.  Again, the light curve is smoother, exhibiting less
structure. Beyond the initial rise in total flux the EVPA exhibits a smooth
variation spanning only tens of degrees, not reproducing the {\it range} in EVPA during the rapid swings.

\section{Discussion of Results from the Modeling}

In Table~\ref{tbl-6} we compare the parameters found for the three sources investigated in this study. The procedures followed allow us to constrain a number of important physical parameters specifying the jet flows of $\gamma$-ray-bright sources  during flaring and to verify that the spectral evolution of the total flux density and of the linear polarization are consistent with the expected spectral evolution for our proposed model incorporating propagating shocks. While an idealized model tracing the time evolution of the emission from an adiabatically-expanding cloud of relativistic particles \citep{laa66}, modified to include a homogeneous magnetic field, was able to predict the main features of the linear polarization variability \citep{all70}, our original shock model \citep{all85a} was developed in an attempt to explain the spectral evolution of polarized radio outbursts in a physically meaningful way. We emphasize that it is  the} spectral evolution of the linear polarization during these outbursts which is the best supporting evidence for the shock scenario which we propose;} such shocks are an often-invoked means to accelerate particles to the high energies required for the production of the $\gamma$-ray flares. While for two of the sources, these shocks are found to be transverse to the flow direction, we do not believe that shocks with this orientation are more prevalent, but rather that this is a selection effect: in the transverse shock case the individual flares in an event are better-resolved and hence more easily modeled. In all three of the events studied the shocks are found to be forward-moving; this result is in contrast to our earlier shock modeling in the mid 1980s and early 1990s where the shocks were found to be reverse \citep{hug91}. While we initially allowed the shocks to be reverse as well as forward, the case of reverse shocks led to values of the Doppler boost and apparent motion which are in disagreement with the excellent VLBI imaging data now available. To generate simulations in agreement with the UMRAO and VLBI results required forward shocks in all three sources.

While there are many free parameters in the simulations, they are, in fact, well-constrained by using the combination of the spectral changes in the total flux density and in the linear polarization.  The simulated linear polarization light curves are highly sensitive to the choice of the observer's viewing angle for fast flows seen at orientations of only a few degrees to the flow direction, and hence linear polarization observations are a powerful constraint on the viewing angle and a way to determine  this important parameter independent of VLBI measurements. We note that the VLBI results commonly quoted are sensitive to the segment of the flow dominating the emission at the observing frequency; hence this method is subject to an inherent bias both because the VLBI observations are sensitive to components which only fill part of the jet flow and  because different frequencies probe different spatial domains.  These differences may also account in part for the differences between the VLBA results in Table~\ref{tbl-1} and our model-derived Lorentz factors: the latter are factors of two to three smaller.

The goal of the work presented here was to verify that the general features of the spectral evolution of the linear polarization and total flux density during the large outbursts selected for detailed analysis can be reproduced by simulations incorporating the adopted shock scenario and not to maximize the fits to selected events. In view of the many simplifications in the model itself, the fact that the simulations are able to reproduce the general data features is very encouraging in supporting this generic model. However, in order to reproduce the details of the light curves, a number of refinements will be required in future work allowing for more realistic models. In particular, we have not investigated the effects of curvature, a common property of AGN jets, on the emission. While swings of 90$^{\circ}$ are plausibly associated with transverse shocks, and swings through smaller ranges can be accounted for by shock obliquity, there are many examples in the UMRAO data of swings through more than 90$^{\circ}$ during linear polarization flares. These are not well-explained by either our simple shock scenario or by stochastic rotator events but might be accommodated by a model where the orientation of the emitting region relative to the observer changes during the evolution of the flare. Additionally we have neglected the changes in the physical properties, most notably density and  magnetic field degree of order, produced by the passage of each shock on subsequent ones. Further we have assumed that all shocks in a given source have the same orientation relative to the flow direction, while individual shocks, more realistically, can develop at any orientation to it \citep{hug05}. The impact of some of these factors on the evolving flow is already being investigated in complementary instability studies \citep{miz14b}. 

Although we have adopted a scenario in which the shocks most likely arise from kinetic Kelvin-Helmholtz instabilities, a variety of other types of instabilities and instability-driven processes have been proposed both globally and microscopically, e.g. \citet{har13}, including those within an MHD scenario in which an ordered magnetic field plays a dominant role \citep{miz14a}.  Current-driven magnetic reconnection has also been proposed, but we are not aware of detailed computations which could be compared with our data. \citet{lyu03} asserts that the same radiation models will hold in both the shock and the reconnection scenarios, but points out that while a power law distribution of the accelerated particles is predicted in both cases, a distinction based on results for a relativistic electron-positron gas is that the spectrum of the accelerated particles is harder in the reconnection case \citep{lar03}. This mechanism has been invoked for explaining  fast flaring observed in the $\gamma$-ray regime \citep{gia13,naj12} in more recent work, but these studies have included only rough estimates of the energetics and no detailed calculations of the emission properties. The low degrees of linear polarization and systematic increases during large outbursts which we find in hundreds of sources monitored over decades has argued for a turbulent magnetic field in the parsec scale domain probed by the UMRAO data and our adopted shock scenario.

The temporally-correlated $\gamma$-ray and radio-band flaring which we find provides support for the interpretation of a parsec-scale origin for at least some $\gamma$-ray flares and  supporting evidence that a single disturbance produces both the $\gamma$-ray and radio-band flares. This general picture agrees with mounting VLBI imaging results in  both the millimeter and centimeter band. The fact that the emission arises in the same spatial region allows us to use the radio-band data to identify the physical conditions at or near to the $\gamma$-ray site. Only a few source events have been modeled to date, but this is a promising way to identify unique conditions associated with the production of the $\gamma$-ray flares detected by {\it Fermi}. Preliminary work comparing the number of shocks associated with paired $\gamma$-ray radio-band flares and orphan radio-band flares (no associated $\gamma$-ray flaring) during nearly identical events in 1156+295  in terms of amplitude and spectral evolution has already identified that complex shock structure and possible shock interactions may play an important role in the production of the high energy events \citep{all14}.

An  important parameter identified from the modeling which impacts emission model studies is the low energy cutoff of the energy distribution of the radiating particles. In the case of OJ~287, this was found to be lower than for the other two sources analyzed. 

The identification of an ordered axial magnetic field in the sources modeled is an unexpected result. The geometry of this ordered fraction of the magnetic field could be either axial or a low pitch angle helix, and  we cannot distinguish between these scenarios. Future modeling including additional events in these and other sources will identify whether this is a common and distinguishing property of $\gamma$-ray-flaring sources.

While we have used independently-determined VLBI results to give initial estimates of some free parameters and for consistency checks and have selectively chosen sources for which there is abundant complementary VLBI data during the tests of our procedures presented here, our method does not depend upon the availability of such data and can be used to determine parameters for $\gamma$-ray-bright sources which have not been well-studied using VLBI techniques. Hence crucial source parameters can be determined for sources which have not 
 been observed intensively with the VLBA but for which single-dish, multifrequency, radio-band polarimetry data are available now or will be available in the future. The use of multi-frequency, radio-brand polarimetry data as the primary constraint in our analysis emphasizes the importance of obtaining such polarimetry data in future programs and their relevance to understanding the broadband blazar phenomenon.

\section{Conclusions}
Our main findings are as follows:
\newline 1. Several centimeter-band events associated with GeV $\gamma$-ray flares detected by the {\it Fermi} LAT exhibit variations in multi-frequency linear polarization and in total flux density expected from propagating shocks internal to the jet flow. 
\newline 2. By comparing simulations incorporating propagating shock structures with the UMRAO data we have constrained the jet flow conditions during three events. In all three cases the shocks are forward-moving with respect to the underlying flow; and the shocks are preferentially oriented transversely to the flow direction.
\newline 3. Inclusion of an underlying ordered component of the magnetic field in the axial direction substantially improves the ability of the simulation to reproduce the observed features in the data.
\newline 4. The amplitude of the fractional linear polarization is very sensitive to the assumed value of the observer's viewing angle. The derived jet directions are nearly line-of-sight and in agreement with independently-determined values from VLBI studies.
\newline 5. The deduced apparent motions, given the viewing angle and the derived shock speed, agree with the speeds determined independently from measurements of
 VLBI-scale component motions.
\newline 6. In one case, OJ 287, an unusually low value of the low energy cutoff of the particle energy distribution is identified.
%% The \notetoeditor{TEXT} command allows the author to communicate
%% information to the copy editor.  This information will appear as a
%% footnote on the printed copy for the manuscript style file.  Nothing will
%% appear on the printed copy if the preprint or
%% preprint2 style files are used.


%% Included in this acknowledgments section are examples of the
%% AASTeX hypertext markup commands. Use \url without the optional [HREF]
%% argument when you want to print the url directly in the text. Otherwise,
%% use either \url or \anchor, with the HREF as the first argument and the
%% text to be printed in the second.

\acknowledgments
We gratefully thank I. Troitskiy and S. Jorstad for sharing data for 0420-014 prior to publication.  We thank the referee for comments which sharpened the presentation and improved the clarity of the paper. This research was supported in part by NASA Fermi Guest Investigator awards NNX09AU16G, NNX10AP16G, NNX11AO13G, and NNX13AP18G, and by a series of grants from the NSF, most recently  AST-0607523 which made the long-term UMRAO program possible. Additional support for the operation of UMRAO was provided by the University of Michigan. T. Hovatta was supported in part by a grant from the Jenny and Antti Wihuri foundation and by the Academy of Finland project number 267324. This research has made use of data from the MOJAVE database that is maintained by the MOJAVE team \citep{lis09}.

%% in parentheses, after the keyword, but they will not be verified.

{\it Facilities:} \facility{UMRAO}, \facility{NRAO}, \facility{Fermi}.

\clearpage

%% the file, uncomment the \plotone, \plottwo, and
%% If you need a layout that cannot be achieved with \plotone or
%% \plottwo, you can invoke the graphicx package directly with the
%% \includegraphics command or use \plotfiddle. For more information,
%% please see the tutorial on "Using Electronic Art with AASTeX" in the
%% documentation section at the AASTeX Web site,
%% http://www.journals.uchicago.edu/AAS/AASTeX.
%%
%% The examples below also include sample markup for submission of
%% supplemental electronic materials. As always, be sure to check
%% the instructions to authors for the journal you are submitting to
%% for specific submissions guidelines as they vary from

%% This example uses \plotone to include an EPS file scaled to
%% 80% of its natural size with \epsscale. Its caption
%% has been written to indicate that additional figure parts will be
%% available in the electronic journal.

\clearpage 

\clearpage

\clearpage 
\begin{figure}
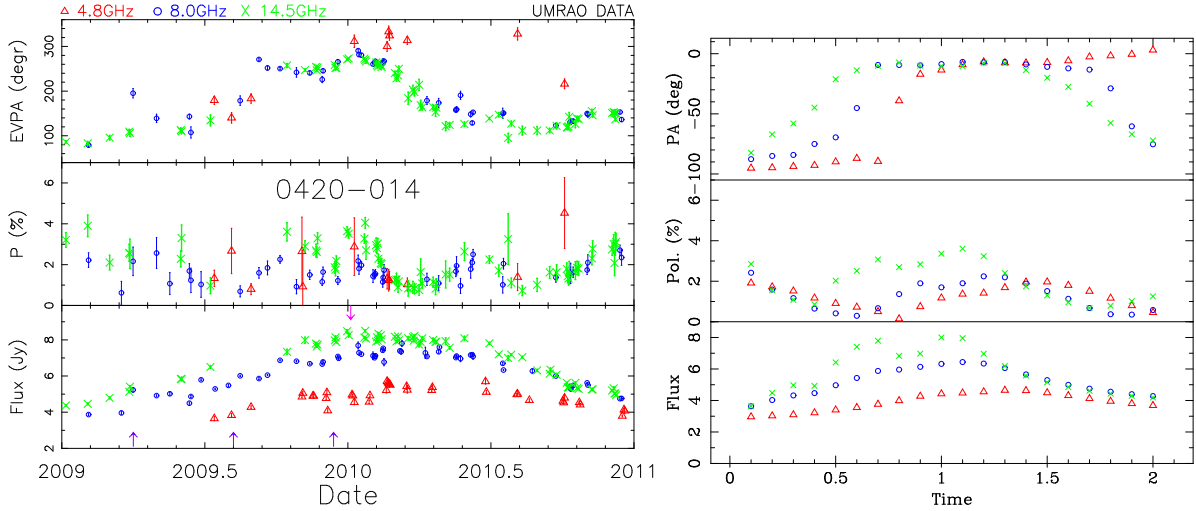

\begin{center}
\includegraphics[scale=0.48]{Fig4a.ps}
\includegraphics[scale=0.33]{Fig4b.eps}
\caption{Comparison of the data and the simulation for the 2009 - 2010 event in 0420-014. Left: daily averages of the total flux density, fractional linear polarization, and EVPA. 
Upward arrows along the time axis mark the shock start times. A downward arrow at the top of the lower panel marks the time of peak $\gamma$-ray photon flux. Right: simulated light curves. The computations have been carried out at 3 harmonically-related frequencies separated by $\sqrt{3}$ which correspond to the UMRAO observing frequencies of 14.5, 8.0, and 4.8 GHz; the symbols follow the convention used
for plotting the UMRAO data.  \label{fig-4}}
\end{center}
\end{figure}

\clearpage

\begin{figure}
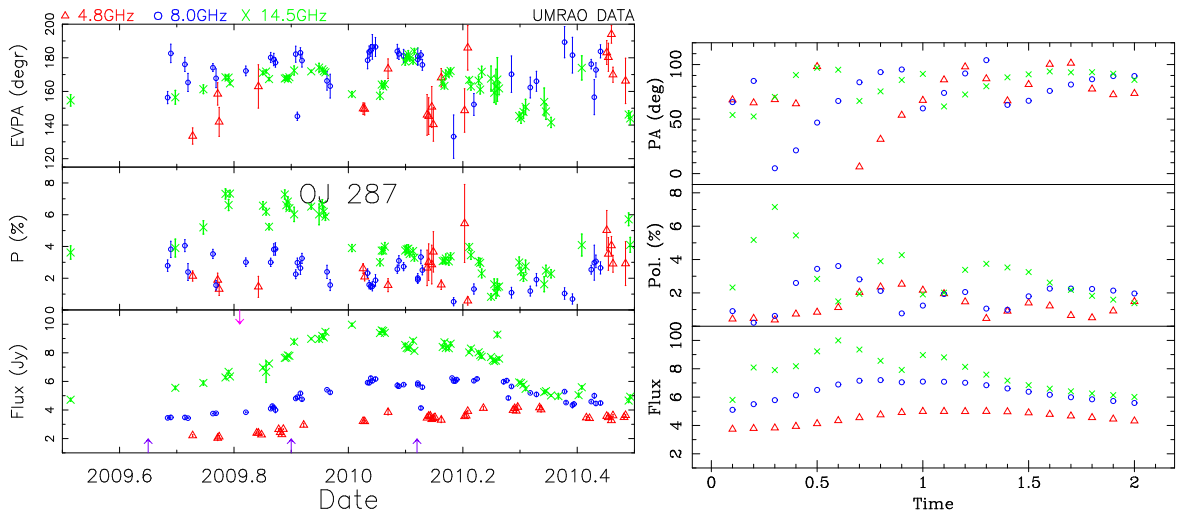

\begin{center}
\includegraphics[scale=0.48]{Fig5a.ps}
\includegraphics[scale=0.33]{Fig5b.eps}
\caption{Comparison of the data and the simulated light curves for OJ 287 during the ten-month event modeled. Details are as in Figure~\ref{fig-4}. Self-absorption produces frequency-dependent time delays in this source and in 0420-014. \label{fig-5}}
\end{center}
\end{figure}

\clearpage

\begin{figure}
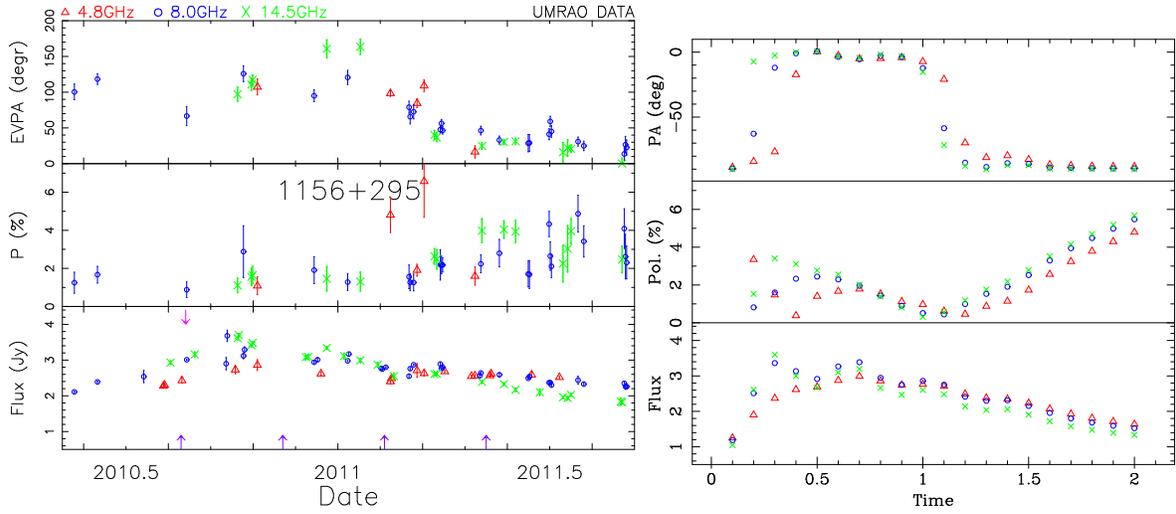

\begin{center}
\includegraphics[scale=0.48]{Fig6a.ps}
\includegraphics[scale=0.33]{Fig6b.eps}
\caption{Comparison of the data and the simulation for the outburst analyzed in 1156+295. The simulation incorporates 4 transverse propagating shocks. Details are as in  Figure~\ref{fig-4}. 
\label{fig-6}}
\end{center}
\end{figure}

\clearpage
\begin{figure}
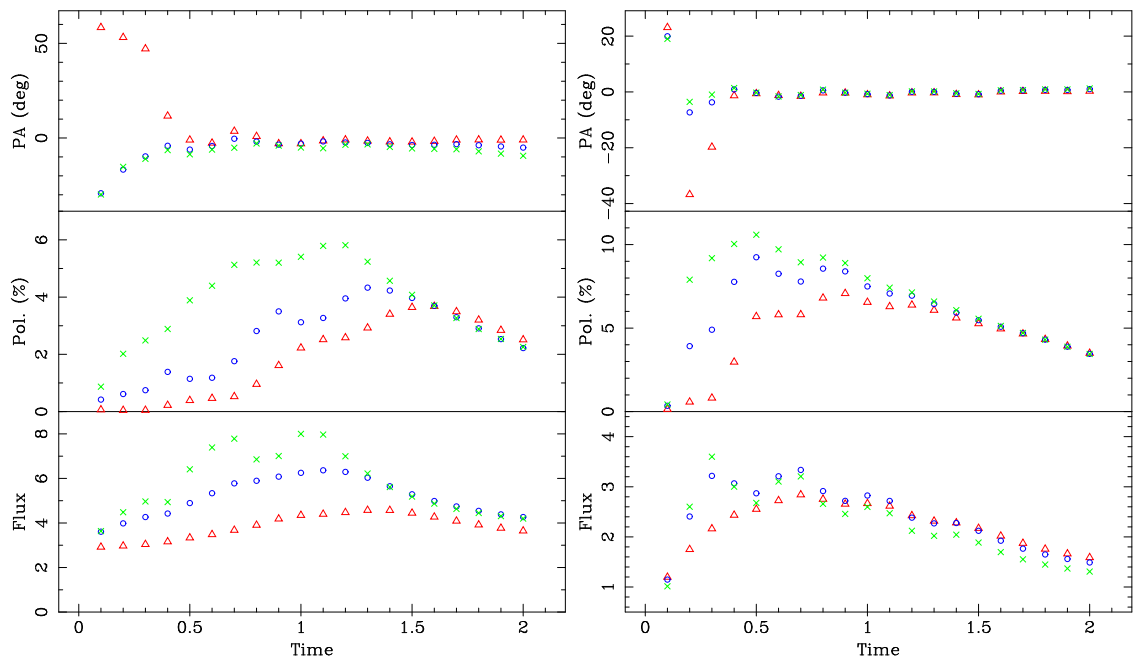

\begin{center}
\includegraphics[scale=0.40]{Fig7a.eps}
\includegraphics[scale=0.40]{Fig7b.eps}
%\plotone{1156Z0bw.eps}
\caption{Model light curves for 0420-014 (left) and for 1156+295 (right) assuming the weakest axial B field which is able to produce a stable EVPA during quiescent states. Comparison with the data for these sources shown in  Figure~\ref{fig-4} (left) and Figure~\ref{fig-6} (left) respectively shows that they do not reproduce the observed features in the linear polarization.
\label{fig-7}}
\end{center}
\end{figure}

\clearpage

\begin{figure}
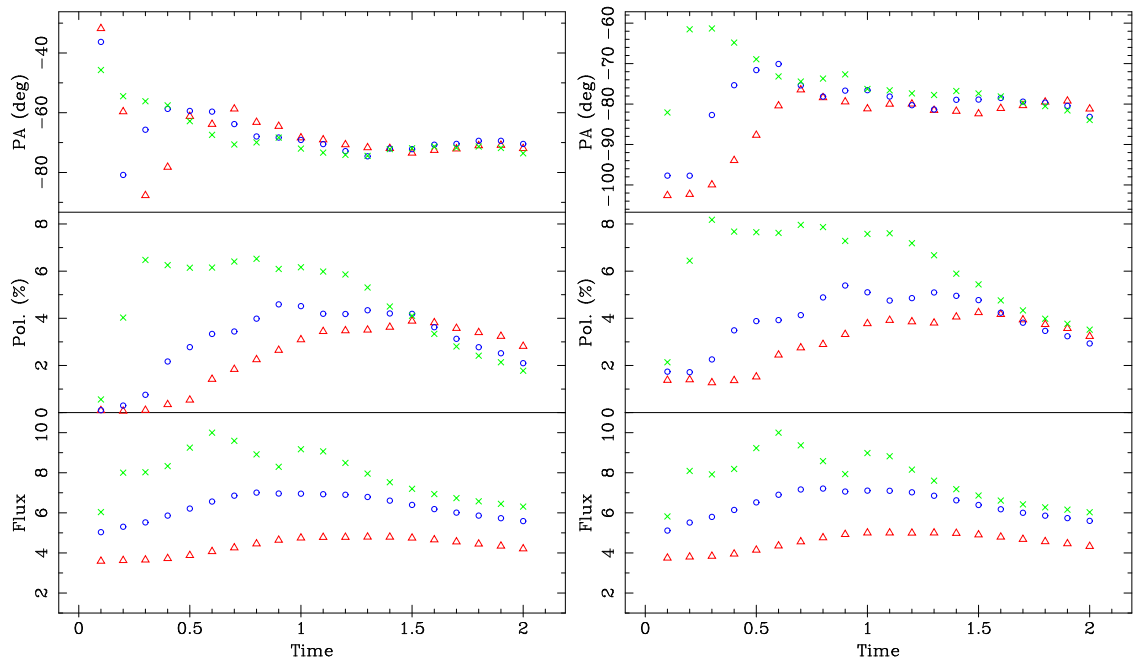

\begin{center}
\includegraphics[scale=0.40]{Fig8a.eps}
\includegraphics[scale=0.40]{Fig8b.eps}

\caption{ Example light curves illustrating the effect of modified input parameters for OJ~287 for comparison with  Figure~\ref{fig-5}. Model light curves for OJ~287 assuming the weakest axial B field which is able to produce a stable EVPA during quiescent states (left) and using the adopted model except for a change in the low energy cutoff of the emitting particles' energy distribution from 20 to 50 (right) are shown. In each simulation all other input parameters are as tabulated.
\label{fig-8}}
\end{center}
\end{figure}

\clearpage

\begin{deluxetable}{lrrrllll}
\tablecaption{Jet Flow Parameters from Centimeter and Millimeter Band Variability\label{tbl-1}}
\tablewidth{0pt}
\tablehead{
\colhead{name}& \colhead{$\beta_{max}$} & \colhead{$\delta_{var}$} &  \colhead{$\Gamma$}
 & \colhead{$\theta_{var}^{\circ}$} &\colhead{$\delta_{43}$} & \colhead{$\Gamma_{43}$} &   \colhead{$\theta_{43}^{\circ}$} 
}
\startdata
0420$-$014 &   5.74c  & 19.9 &  10.8  &  1.5 & 16.2$\pm$1.2 &  11.0$\pm$0.5 &  3.0$\pm$0.6\\
OJ~287     &   15.14c & 17.0 &  15.3  &  3.4 & 13.6$\pm$2.4 & 16.3$\pm$3.8 &  4.0$\pm$0.6\\
1156$+$295 &   24.58c & 28.5 &  24.9  &  2.0 & \nodata      & \nodata      &  \nodata    \\
\enddata
\tablerefs{$\beta_{max}$ from the MOJAVE website; variability Doppler factor and viewing angle as described in text; 43 GHz Doppler factor, Lorentz factor, and viewing angle from \citet{jor05}.}
\end{deluxetable}

\clearpage

\begin{deluxetable}{lccc}
%\tabletypesize{\scriptsize}
%\rotate
\tablecaption  {Parameters for Individual Shocks: 0420-014 \label{tbl-2}}
\tablewidth{0pt}
\tablehead{
\colhead{Shock} & \colhead{1}   & \colhead{2}   & \colhead{3}  \\ 
}
\startdata
     start (t$_0$)        & 2009.25  & 2009.6 & 2009.95 \\
    length (l)            & 10.0  &  15.0   &  10.0    \\
    compression ($\kappa$) & 0.8   &  0.66  &  0.65     \\
location of S$_{max}$      & 0.22   & 0.64   & 1.06   \\
\enddata

\tablecomments{The length of the shocked flow is expressed as a percentage of the flow length. The location of the peak flux is given as the fraction of the time window. This convention is adopted in the following tables.}
\end{deluxetable}

\begin{deluxetable}{lcccc}
%\tabletypesize{\scriptsize}
%\rotate
\tablecaption  {Comparison of shocks with 43 GHz VLBI components: 0420-014 \label{tbl-3}}
\tablewidth{0pt}
\tablehead{
\colhead{Shock} & \colhead{1}   & \colhead{2}   & \colhead{3}    &  \colhead{}  \\ 
}
\startdata
     Shock start (t$_0$)       & 2009.25  & 2009.6   & 2009.95    &  \nodata  \\
     nearest component (ID)    &  K2    &  K3      &  K4          &   K5    \\
     T$_{eject}$ of component    & 2008.72*  & 2009.444$\pm$0.110 &  2009.892$\pm$030  &   2010.247$\pm$0.004 \\
%     T of PeakS$_K$ JD    & 55055    & 55120     &  55245   &   55510    \\   \nodata
%    T of PeakS$_K$ Greg  & 2009.65  & 2009.79  & 2010.13  &  2010.86    \\
     $\beta_{app}$ (component)  & 13.43$\pm$3.35c       &  16.58$\pm$1.39c      & 13.08$\pm$1.48c   &  25.30$\pm$0.49c  \\
     $\beta_{app}$ (shock)      &  11c    &  11c    &  11c      &   \nodata          \\
    \enddata

\tablerefs{The VLBI values tabulated are from \citet{tro13}. * indicates a private communication from S. Jorstad \& I. Troitskiy.}
\end{deluxetable}

\clearpage

\begin{deluxetable}{cccc}
%\tabletypesize{\scriptsize}
%\rotate
\tablecaption  {Parameters for Individual Shocks: OJ~287 \label{tbl-4}}
\tablewidth{0pt}
\tablehead{
\colhead{Shock} & \colhead{1}   & \colhead{2}   & \colhead{3}  \\ 
}
\startdata
    start (t$_0$)          & 2009.65 & 2009.9 & 2010.12 \\
    length (l)             & 2.5     &  2.5   &  6.0    \\
    compression ($\kappa$) & 0.5     &  0.7   &  0.7     \\
  location of S$_{max}$ & 0.20       & 0.6    & 1.05   \\
\enddata

\end{deluxetable}

\clearpage

\begin{deluxetable}{lcccc}
%\tabletypesize{\scriptsize}
%\rotate
\tablecaption   {Parameters for Individual Shocks: 1156+295 \label{tbl-5}}
\tablewidth{0pt}
\tablehead{
\colhead{Shock} & \colhead{1}   & \colhead{2}   & \colhead{3}   & \colhead{4} \\ 
}
\startdata
    start (t$_0$)         & 2010.63  & 2010.87 & 2011.11 & 2011.35 \\
    length (l)            & 10.0     &  10.0   &  10.0 & 10.0  \\
    compression ($\kappa$) & 0.5     &  0.6   &  0.7  & 0.8   \\
location of S$_{max}$      & 0.016    & 0.64    & 0.94 & 1.36   \\
\enddata

\end{deluxetable}

\clearpage

\begin{deluxetable}{lccc}
%\tabletypesize{\scriptsize}
%\rotate
\tablecaption  {Summary of Model Parameters \label{tbl-6}}
\tablewidth{0pt}
\tablehead{
\colhead{Parameter} & \colhead{0420-014}   & \colhead{OJ 287} & \colhead{1156+295}    
 }
\startdata 
  Spectral Index ($\alpha$)  & 0.25   &  0.25   & 0.25 \\
  Fiducial Lorentz Factor ($\gamma_{c}$)   & 1000    &  1000  &   1000   \\
  Cutoff Lorentz Factor ($\gamma_{i}$)     &  50     &    20  &   50   \\
  Bulk Lorentz Factor       &  5.0    &   5.0 &   10   \\
 Number of Shocks           & 3       &  3     &   4   \\
 Shock obliquity            &  90$^{\circ}$ &  30$^{\circ}$ & 90$^{\circ}$ \\
  Shock Sense                     & F       &     F  &  F   \\ 
 Viewing angle ($\theta_{obs}$)   &   4$^{\circ}$  & 1.5$^{\circ}$ &  2.0$^{\circ}$   \\
$\beta_{app}$                &  11c             &  17c          &  22c              \\
Axial Magnetic Field*         &  16\%             &   50\%        & 50\%            \\
\enddata

\tablecomments{The axial magnetic field is given in units of the total magnetic field energy density.}
\end{deluxetable}

\end{document}